\documentclass{ws-ijmpe}
\usepackage{units}

\newcommand{\pt}{p_\perp}
\newcommand{\kt}{k_\perp}

\newcommand{\mean}[1]{\langle #1 \rangle}
\newcommand{\etal}{{\it et al.}}

\begin{document}

\markboth{K. Zapp, G. Ingelman, J. Rathsman, J. Stachel}{Heavy Quark Energy Loss Through Soft QCD Scattering in the QGP}

\catchline{}{}{}{}{}

\title{HEAVY QUARK ENERGY LOSS\\ THROUGH SOFT QCD SCATTERING IN THE QGP}

\author{\footnotesize KORINNA ZAPP}

\address{Physikalisches Institut, Universit\"at Heidelberg, Philosophenweg 12\\
D-69120 Heidelberg,
Germany\\
zapp@physi.uni-heidelberg.de}

\author{GUNNAR INGELMAN}

\address{High Energy Physics, Uppsala University, Box 535\\
S-75121 Uppsala, Sweden\\
gunnar.ingelman@tsl.uu.se}

\author{JOHAN RATHSMAN}

\address{High Energy Physics, Uppsala University, Box 535\\
S-75121 Uppsala, Sweden\\
johan.rathsman@tsl.uu.se}

\author{\footnotesize JOHANNA STACHEL}

\address{Physikalisches Institut, Universit\"at Heidelberg, Philosophenweg 12\\
D-69120 Heidelberg,
Germany\\
stachel@physi.uni-heidelberg.de}

\maketitle

\begin{history}
\received{(received date)}
\revised{(revised date)}
\end{history}

\begin{abstract}
A strong suppression of non-photonic electrons in Au+Au collisions is observed at \textsc{Rhic}. This is in disagreement with the expected dominance of the energy loss via gluon radiation, which predicts a much weaker suppression of heavy flavours due to the dead cone effect. However, collisional energy loss is also important, as demonstrated recently by the Soft Colour Interaction Jet Quenching model. Based on this model we show that collisional energy loss of charm quarks gives a suppression of electrons of the observed magnitude, but the contribution from beauty decays, which dominates at intermediate and large $\pt$, results in a somewhat weaker overall suppression of electrons than observed.
\end{abstract}

\section{Introduction}

In deep inelastic $ep$ and $pp$ scattering hard scattered quarks (and gluons) have to traverse the proton remnant, since the hard interaction takes place inside the proton. It cannot be expected, that interactions cease below the perturbative cut-off, where the coupling is large, but typical momentum transfers are small. However, it was found\cite{sci}, that such non-perturbative interactions may be important. As the hard parton travels through the coloured background of the proton remnant, it may interact with it and although these soft interactions do not change the kinematics, they do change the colour topology. This can lead, for example, to diffractive events with large rapidity gaps. This circumstance is modeled by soft gluon exchanges in the successful phenomenological SCI (Soft Colour Interaction) model\cite{sci}, but can also be understood theoretically in terms of multiple gluon exchange as described by the Wilson line in the parton density definition\cite{sci-theo}.

Following the ideas behind the SCI model it seems plausible that if soft scattering is important in DIS, it may also be important for a hard-scattered quark or gluon traversing a quark-gluon plasma (QGP). In the QGP many more coloured scattering centres are available, so that even a small momentum transfer per scattering adds up to a sizeable effect, since the interaction probability is large so that one parton may experience many interactions.

\section{The SCI Jet Quenching Model}

The SCI Jet Quenching model\cite{karo} has emerged from the SCI model and describes the rescattering of an energetic quark or gluon via soft interactions in a QGP.

The QGP is represented by an ideal relativistic gluon gas (quarks are neglected because the plasma is initially gluon rich and the gluons come faster into thermal equilibrium). The gluon and energy density are accordingly given by  
$n_\text{g} = g_\text{g}\, T^3 2\,\zeta(3)/2\pi^2$ and $\epsilon_\text{g} = \pi^2 g_\text{g}\,T^4/30$ respectively ($g_\text{g} = 16$ is the degeneracy).
The plasma expands longitudinally (Bjorken expansion\cite{bjorken}), so that the time dependence of the gluon density and temperature is given by $n_\text{g}(\tau) \propto \tau^{-1}$ and $T(\tau) \propto \tau^{-1/3}$, where $\tau = \sqrt{t^2-z^2}$ is the proper time. This rapid decrease of the gluon density plays an important role for the phenomenology of the model. As initial condition the energy density is fixed at $t_0 = \unit[1]{fm/c}$, although the plasma formation time can be shorter than that ($\tau = 0$ refers to the instant of maximum overlap of the nuclei). The energy density profile is proportional to the number of participating nucleons per transverse area $n_\text{part}(x,y)$, since the initial energy production is presumably dominated by soft processes and should therefore scale with the number of participants. The parameter $\epsilon_0$ ($\unit[5.5]{GeV\,fm^{-3}}$) gives the mean energy density for a central ($b=0$) collision at $\tau = \unit[1]{fm/c}$, this fixes the normalisation of the energy density distribution for all centralities.

The number of binary collisions and participants and the geometrical cross section as function of the impact parameter are calculated in the framework of a simple Glauber model\cite{eskola} using different potentials (sharp sphere and Woods-Saxon). 

The nucleons may build up transverse momentum in soft scatterings prior to the hard interaction (Cronin effect). This is included by increasing the width $\sigma_{\kt}^2(x,y,b) = \sigma_{\kt,0}^2 + \alpha\cdot (N_\mathrm{scat}(x,y,b)-1)$ of the Gaussian primordial $\kt$-distribution of the partons by an amount $\alpha$ for each soft rescattering\cite{cronin}. The parameter $\alpha=\unit[0.5]{GeV^2}$ is obtained by fitting d+Au data \textsc{Rhic} and does not depend on the hard scale, since parton showers are treated explicitly.

The perturbative part and the hadronisation is simulated using \textsc{Pythia\,6.4}\cite{pythia}. 
The hard partons are produced from LO matrix elements (for heavy quarks massive matrix elements with $m_\text{c}=\unit[1.5]{GeV},m_\text{b}=\unit[4.8]{GeV}$ are used) and initial and final state parton showers are simulated. Each of the partons is then tracked through the QGP and may interact with it. The interactions start at the plasma formation time $\tau_\text{i}$ ($\unit[0.2]{fm/c}$) and stop when the local energy density drops below the critical temperature $T_\text{c}$ ($\unit[175]{MeV}$) or the parton leaves the overlap region. In the QGP a parton interacts with each  plasma-gluon (with effective mass $m_\text{g}=\unit[0.2]{GeV}$) that is closer than a screening length $R_\text{scr}=\unit[0.3]{fm}$ (alternative parameter value $\unit[0.5]{fm}$) with a probability $p_\text{q}=0.5$ for quarks and $p_\text{g}=0.75$ for gluons having a larger colour charge. The interactions are elastic scatterings where the squared momentum transfer $t$ has a Gaussian distribution with width $\sigma_t=\unit[0.5]{GeV^2}$. We do not use perturbative QCD matrix elements since the typical momentum transfer is a few hundred \unit{MeV}, i.e.\ in the non-perturbative regime. Depending on the value of the screening length parameter $R_\text{scr}$ being $\unit[0.3]{fm}$ or $\unit[0.5]{fm}$, this results in a soft scattering cross section $\sigma^\text{sci}=\unit[1.9]{mb}$ or $\unit[5.2]{mb}$. 

After leaving the overlap region the partons are hadronised using independent fragmentation\cite{indep}, since the colour connections among the partons produced in the perturbative phase are destroyed. A more detailed discussion of the model is available\cite{karo} and we note that the parameters are the same for light and for heavy flavours.

\section{Results}

The results are shown for Au+Au collisions at $\sqrt{s^\text{NN}} = \unit[200]{GeV}$ at mid-rapidity ($|\eta| < 0.35$). A p+p reference is simulated within the same framework and is in reasonable agreement with measurements. Charm and beauty are simulated separately and added according to the production cross sections. 

\begin{figure}[th]
\centerline{\input{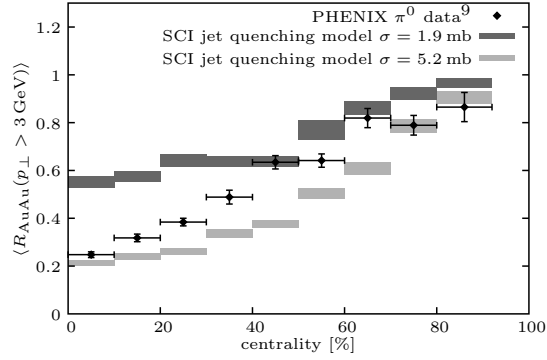}}
\vspace*{8pt}
\caption{Nuclear modification factor for neutral pions averaged between \unit[3]{GeV} and \unit[8]{GeV} transverse momentum as function of the centrality of the collision (fraction of geometric cross section), \textsc{Phenix} data and SCI Jet Quenching Model (Woods-Saxon potential) with two different scattering cross sections.}
\label{fig_light}
\end{figure}

A summary of the model results\cite{karo} for the nuclear modification factor $R_\text{AuAu}$ for neutral pions in shown in Fig.\ \ref{fig_light}. With the small scattering cross section the SCI Jet Quenching model can account for roughly \unit[50]{\%} of the observed $\pi^0$ suppression in central collisions and comes closer to data for more peripheral events. With the increased cross section it is in agreement with the data for peripheral and very central collisions, but the increase with centrality is not quite linear as in the data. $R_\text{AuAu}$ increases weakly with transverse momentum $\pt$, see details in our publication\cite{karo}.

\begin{figure}[th]
\centerline{\input{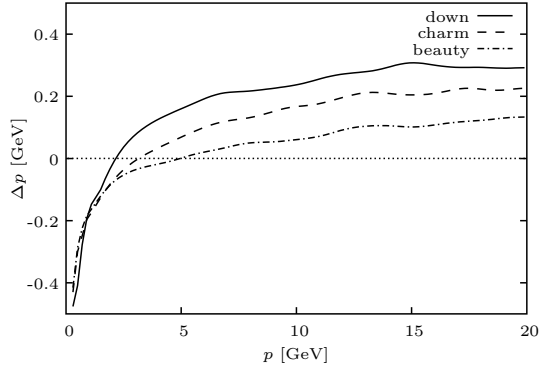}}
\vspace*{8pt}
\caption{Momentum loss in the first scattering of different quark flavours versus their momentum.}
\label{fig_eloss}
\end{figure}

Fig.\ \ref{fig_eloss} shows a comparison of the momentum loss in a single scattering for different quark species. The heavier the quark is the less energy (or momentum) it looses. This is simply due to kinematics, since a heavy object does not take as much recoil as a light one. The negative part of $\Delta p$ shows that low energy partons may gain momentum if their energy is smaller than the thermal energy of the surroundings (Fig.\ \ref{fig_eloss} shows the first scattering, so the temperature is quite high).

\begin{figure}[th]
\centerline{\input{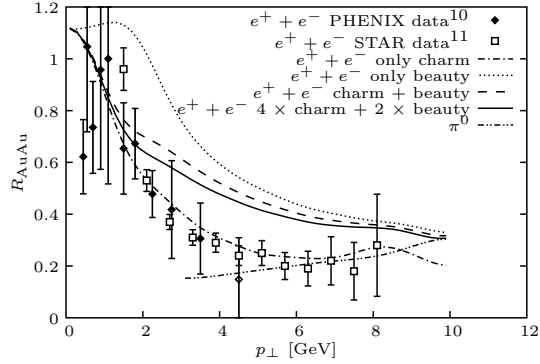}}
\vspace*{8pt}
\caption{Nuclear modification factor of electrons from heavy flavour decays in central collisions, model calculations ($\sigma^\text{sci} = \unit[5.2]{mb}$, Woods-Saxon potential) are shown for electrons from charm and beauty decays separately as well as the sum: dashed line for charm and beauty having the same K-factor, solid line for charm K-factor twice the beauty one. The model result for the nuclear modification factor of neutral pions is also shown for comparison.}
\label{fig_heavy}
\end{figure}

In Fig.\ \ref{fig_heavy} we show the results for the suppression of electrons (and positrons) from heavy flavour decays. While the model gives the right suppression of $\pi^0$, with the same parameters the electron suppression is somewhat weaker than measured by the experiments. The beauty suppression is clearly weaker than the charm suppression as one would na\"ivly expect because of the smaller energy loss. Since the electron spectrum at intermediate and high $\pt$ is dominated by the beauty contribution, the sum is relatively close to the beauty result. 
 There is a substantial uncertainty on the K-factor for charm and beauty, but in this ratio only their difference is important. Given that pQCD calculations indicate that the charm K-factor may be up to twice the one for beauty\cite{cacciari}, Fig.\ \ref{fig_heavy} show the results for K-factor ratio of unity or two. Fortunately, the result depends only weakly in these differences between the K-factors, since at large $\pt$ the sum is so much dominated by the beauty contribution.

\begin{figure}[th]
\centerline{\input{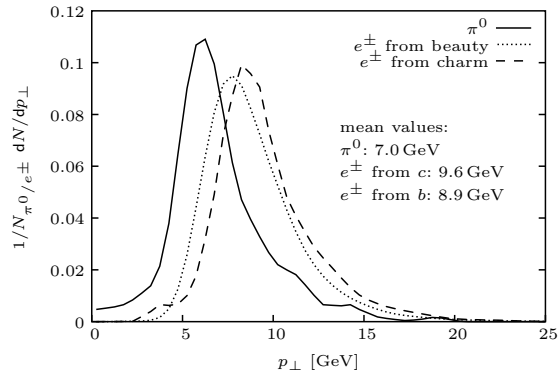}}
\vspace*{8pt}
\caption{$\pt$ of quarks that fragment into a $\pi^0$ with $\pt = \unit[5 \pm 0.25]{GeV}$ (light flavours including gluons) or into a $e^\pm$ with $\pt = \unit[5 \pm 0.25]{GeV}$ (heavy flavours).}
\label{fig_parent}
\end{figure}

The nuclear modification factors of electrons from charm decays and neutral pions are similar, in spite of the smaller energy loss of charm quarks. Moreover, the slope of the $\pt$ spectrum of charm and the light flavour cocktail is nearly the same at intermediate and high $\pt$. Fragmentation and decay affect $R_\text{AuAu}$; pions at $\pt = \unit[5]{GeV}$ stem from a quark or gluon with on average $\unit[7]{GeV}$ transverse momentum and electrons of the same $\pt$ typically originate from a charm quark of a higher transverse momentum of \unit[9.6]{GeV}, see Fig.\ \ref{fig_parent}. Together with the steeply falling $\pt$ spectrum this can explain the similar suppression of $\pi^0$ and electrons from charm decays. The beauty parent quarks lie between the other two, but the slope of the $\pt$ spectrum is again similar, so that the suppression of electrons from beauty decays is weaker. 


\section{Conclusions}

In order to clarify the question, to what extent energy loss through scattering can help to understand the strong suppression of electrons from heavy flavour decays, we present results from the SCI Jet Quenching model with parameters chosen such that it reproduces the full $\pi^0$ suppression. 
 The main effect in the observed suppression of electrons can then be accounted for, but the model does not give large enough suppression at large $\pt$. This is due to the weak suppression of  electrons from beauty decays, while the suppression of charm-electrons seems sufficient. Thus, a pure collisional energy loss model cannot account for the full electron suppression. The situation will be worse in models that include also radiative energy loss, which is weaker for heavy flavours. The difference is here even stronger than in our model\cite{djordjevic}. Since the contribution of collisional energy loss would have to be reduced in order to get the correct amount of $\pi^0$ quenching, the suppression of electrons  cannot be stronger than in our model (in fact a similar conclusion is drawn in\cite{wicks}). It seems, therefore, that the conventional energy loss models miss some important part of physics for the heavy flavours.

\end{document}